\documentclass[a4paper,11pt]{article}
\pdfoutput=1 
\usepackage[utf8]{inputenc}
\usepackage{jheppubmod}

\usepackage{graphicx,xcolor,amsmath,amssymb,fancyhdr,siunitx,mathtools}
\usepackage[numbers]{natbib}
\usepackage[titletoc]{appendix}
\numberwithin{equation}{section}

\usepackage{cleveref}

\def\dslash{\not{\hbox{\kern-2pt $\partial$}}}
\def\Dslash{\not{\hbox{\kern-4pt $D$}}}
\def\Oslash{\not{\hbox{\kern-4pt $O$}}}
\def\Aslash{\not{\hbox{\kern-4pt $A$}}}
\def\partialslash{\not{\hbox{\kern-4pt $\partial$}}}
\def\Qslash{\not{\hbox{\kern-4pt $Q$}}}
\def\pslash{\not{\hbox{\kern-2.3pt $p$}}}
\def\kslash{\not{\hbox{\kern-2.3pt $k$}}}
\def\qslash{\not{\hbox{\kern-2.3pt $q$}}}
\def\svslash{\not{\hbox{\kern-2.3pt $sv$}}}
\usepackage{verbatim}

\newcommand{\md}[1]{\textcolor{blue}{[MD: #1]}}

\title{Remarks on the Axion Domain Wall Problem}

\author{Michael Dine,}

\emailAdd{mdine@ucsc.edu}

\affiliation{
    Santa Cruz Institute for Particle Physics and Department of Physics,\\University of California, Santa Cruz,\\
    Santa Cruz, CA, USA
}

\abstract{Theories in which the Peccei-Quinn phase transition occurs after inflation tend to suffer from problematic domain walls.  One possible solution involves a small, explicit breaking
ot the symmetry.  But this raises other potential issues.  We review some aspects of axion domain walls, focussing especially on this proposed solution.  We argue, in disagreement
with some recent literature, that there is little axion radiation from the system until the domains actually collapse.  The same applies to gravitational waves and electromagnetic
radiation.  The final stages of the collapse yields small numbers of extremely energetic axions, which interact only rarely with ordinary matter, and are thus relatively harmless.  We then note that, if one
accepts a remarkable coincidence, this solution can be acceptable.  We consider a possible explanation of the required coincidence.}
\begin{document}
\maketitle

\section{To Do}

Consider domain walls bounded by strings, along the lines of the old paper by Sikivie et al.  Is there any change in the picture, e.g. due to attractive and repulsive forces between string elements?  Might imagine
always have order one domain per horizon before considering bias.  Suppose, e.g., ${\cal N}$.   In light travel time to cross the horizon, system develops a large $\gamma$.  

Ask what fraction of domain wall energy might be in hadrons and interpret in terms of final collapse.

Think about final collapse in terms of particle collisions.  If principally axions, what are the collision products?  

\section{Introduction}

  A Peccei-Quinn symmetry\cite{pq1,pq2, weinbergaxion,wilczekaxion} has the potential to solve the strong CP problem {\it and} account for the dark matter of the universe
 \cite{abbottsikivie,preskillwisewilczek,dinefischleraxiondarkmatter}.  Before considering cosmology, the axion decay constant, a priori,
  can take a broad range of values.  Stellar astrophysics places a lower bound in the range of $10^{9}-10^{10}$ GeV\cite{kolbturner}.  Big bang cosmology, with the assumption
  that the universe, in the past, was hotter than a GeV or so, places an upper limit of about $10^{12}$ GeV.  Attaining a symmetry of sufficient {\it quality}\cite{kmr,axionquality} to solve the strong CP problem, however,
  is quite a challenge.  String theory, and more general considerations of quantum gravity, rule out exact, continuous global symmetries.  So one expects
  that in the effective field theory at low energies, at the very least there will be Planck-suppressed operators which violate the symmetry.  Even for the low range of $f_a$,
  operators of very high dimension must be suppressed to account for the smallness of $\theta$\cite{kmr}. One might try to account for this suppression by discrete symmetries\cite{dinediscretesymmetries},
  but the symmetries must be quite large.  String theory appears capable of avoiding this problem\cite{wittenaxions}, in the sense that PQ symmetries may be violated only by non-perturbative
  effects, which can be extremely small if the theory generates a small coupling constant.  But in this case, a value of $f_a$ much smaller than, say, typical scales associated with coupling constant
  unification, would be surprising.  Larger scales are admissible if the universe was never much hotter than nucleosynthesis temperatures in the past.  This might occur if there was a period
  where the universe was dominated by moduli; see, for example\cite{dinebook}.
  
  These observations arguably cast doubt on the PQ solution, and in any case, would seem to favor large values of $f_a$ and a modified cosmology.   In this paper, however,
  we will adopt the conventional picture that the universe was quite hot in the past and we will assume that there was a PQ transition after inflation, and focus on the problem of domain walls.

In the scenario in which there is a PQ transition after inflation, domain walls are potentially problematic\cite{sikiviedomainwalls}.If the PQ symmetry is  an exact, continuous global symmetry (up to anomalies), the theory has stable domain walls provided the 
coefficient of the QCD anomaly is (suitably normalized) an integer different than one.  The domain wall energy density falls off as $1/R$, where $R$ is the scale factor, as opposed to the radiation energy density ($1/R^4$)
or the matter dominated energy density $1/R^3$.  Typically, the domain walls dominate well before the present era, spoiling the successes of the Standard Cosmology.

Two plausible solutions to this problem were put forward in\cite{sikiviedomainwalls}:
\begin{enumerate}
\item  The coefficient of the anomaly is unity.
\item  There is a small explicit breaking of the PQ symmetry, large enough to lead to collapse of the domain wall system.
\end{enumerate}
Even for the first solution, one has to consider the effects of cosmic strings\cite{sikiviedomainwallstrings}.  The second solution has troubling features.  The strength of the leading
symmetry breaking operator is restricted to a narrow range.  First, it must be small enough that the resulting $\theta$ satisfies the current experimental bounds.  Typically this requires that the leading operator
which breaks the continuous PQ symmetry is of very high dimension\cite{kmr}.  Second, the operator must be {\it large} enough that, if there are domain walls, these disappear before
they come to dominate the energy density of the universe. Naively, for interesting values
of $f_a$, the axion decay constant, these two conditions limit the symmetry breaking operator to a narrow range (we will take $f_a = 10^{12}$ GeV as our benchmark).  If the suppression of high dimension operators is a consequence of discrete symmetries,
these symmetries must be very large, but not {\it too} large\cite{dinediscretesymmetries}.

Recently, the authors of \cite{sarkar} have revisited the domain wall problem.  They put forward and then rule out a third possible solution: a bias in the domain wall ensemble favoring one of the ground
states.They then argue that the second solution above is unlikely to work, except for relatively small values of the axion decay constant, small enough as to be problematic for stellar processes.
They argue, in particular, that the collapsing domain wall system produces too much dark matter.

Motivated by the appearance of high quality PQ symmetries in string theory, the present author has long been an advocate for high scale breaking of the PQ symmetry, which requires a breakdown of
the Standard Big Bang Cosmology at temperatures not much higher than nucleosynthesis temperature.    But in this note, we consider the possibility of a post-inflationary
PQ transition and smaller axion decay constants, with a conventional
thermal history for the university at least up to temperatures of a few GeV.  We will focus on the issues associated with the second solution, small breaking of the Peccei-Quinn symmetry.
We will consider the dark matter issue, demonstrating that, until the domains collapse, most of the excess energy is converted into kinetic energy of the domain walls.  Once the domains shrink to sizes of order
$m_a^{-1}$, this energy is converted to ultrarelativistic axions and hadrons.  Provided that the the domain walls never dominate, these objects are relatively harmless.

We will review the problem of accounting for small symmetry breaking, focussing on models where the PQ symmetry is an accident of a large
discrete symmetry\cite{dinediscretesymmetries}.  Such symmetries, at least at first sight, are not particularly plausible.  The requirements, indeed, yield extremely large symmetries, yet the symmetry
also cannot be too large if the breaking is to be sufficient to avoid a domain wall dominated universe.
Given the seeming absurdity of the requirements on the symmetry breaking, we ask whether there might be some anthropic explanation.  Taking a very conservative approach to the anthropic principle, where one asks
whether the change of one particular parameter can rule out the existence of observers, anthropic constraints have the potential to restrict the strength of the symmetry breaking to the required range.

This note is organized as follows.
In the next section, we review some aspects of domain walls and the cosmic strings which bound them\cite{sikiviedomainwalls}.  In particular, we discuss the sense in which one can systematically
construct the domain wall from the chiral lagrangian, and also provide a simple, analytic domain wall solution in a particular limit.  In section \ref{tilt}, we study the system in the presence of explicit breaking.
We discuss constraints on the size of the breaking and the axion decay constant.  We focus, particularly, on models where the PQ symmetry arises from a large discrete symmetry, noting that the
symmetry must be quite large to accommodate the current limits on $\theta$ but (anticipating our cosmology discussion) can't be appreciably larger than this if it is to avoid domain wall catastrophes.  In section \ref{cosmology} we turn to cosmology.  After reviewing aspects of the cosmological domain wall problem, we demonstrate that the wall collisions principally produce gravitational waves and note that their energy density
can readily be in a suitable range.  We then turn to the coincidence problem in section \ref{anthropic}, arguing that it places requirements on a theory which call out, if a PQ symmetry is realized in nature
in this fashion, for
an anthropic solution.  As noted above,  we will see that such a solution is plausible.  It is hard to see how otherwise there would be any solution at all. Our conclusions are presented in section \ref{conclusions}.

\section{Domain Wall Generalities}

Suppose we have an exact Peccei-Quinn symmetry up to the anomaly.  Under the PQ symmetry, the various fields transform by phases $e^{i q_{pq} \alpha}$.  By convention, we take the PQ charges, $q_{pq}$ to be integers.    $\theta$ changes, in general, by $2 \pi {\cal N}$ for some integer ${\cal N}$ under this transformation.  Given $2\pi$ periodicity of $\theta$, the symmetry is in fact
$Z_{\cal N}$.  For ${\cal N } \ne 1$, there are domain walls.

The tension of the domain walls is of order 
\begin{equation}
T \sim
m_a f_a^2 \sim m_\pi f_\pi f_a.
\end{equation}
We will have in mind $f_a \sim 10^{11}- 10^{12} ~{\rm GeV}$ in what follows.
It is of interest to ask whether, within the framework of chiral perturbation theory and/or large $N$, we can write a strict equality for the domain wall tension.  %We will explore this question in section \ref{largen}.

\subsection{Domain Wall Solutions from the Chiral Lagrangian}
\label{largen}

Just as one can compute the axion mass using the chiral lagrangian, one can obtain the domain wall solutions in the case that the system supports axionic domain walls.
Suppose that the light quarks have PQ charges $q_i$ (we will mainly write explicit formulas for the case of two light quarks).   Suppose, as well, that the PQ symmetry has
anomaly
\begin{equation}
\partial_\mu j^{\mu}_{PQ} ={{\cal N} \over 32 \pi^2} F \tilde F.
\end{equation}  
Then we can define an anomaly free current, $\tilde j^{mu}_{PQ}$ by subtracting off a non-conserved current with the same anomaly:
\begin{equation}
j^{5 \mu} ={\cal N}( \bar u \gamma^\mu \gamma^5 u + \bar d \gamma^\mu \gamma^5 d)
\end{equation}
So now:
  \begin{equation}
  \partial_\mu \tilde j_{PQ}^\mu ={\cal N}( m_u \bar u \gamma^5 u+ m_d \bar d \gamma^5 d).
  \end{equation}
  In computing the axion mass, one sometimes makes a different choice\cite{bardeenetal}, so that the divergence of the current does not have matrix elements between vacuum and the single pion state,
  but this is not convenient for the domain wall problem, where one needs to consider a finite axion field range.
  
  We can explore the effect of finite transformations generated by $\tilde Q_{PQ}$, 
  \begin{equation}
  U(\alpha) = e^{i \alpha \tilde Q_{PQ}}.
  \end{equation}
  Under this transformation the quark mass terms are not invariant; these transform as:
  \begin{equation}
  m_u \bar u u + m_d \bar d d  \rightarrow ( m_u \bar u u + m_d \bar d d) \cos ({\cal N} \alpha) +  (m_u \bar u \gamma_5 u + m_d \bar d \gamma_5 d)\sin({\cal N} \alpha) . 
  \end{equation}
Now consider the chiral lagrangian.  Our goal is to integrate out the pion fields.  We do this by replacing $\bar u u$ and $\bar d d$ by their expectation values as functions of the pseudogoldstone  fields, and solving for the minimum of the $\vec \pi$ potential
  as a function of $a = \alpha f_a$.  Switching to a two component notation, and letting $f,g$ denote flavor indices:
  \begin{equation}
   \bar \psi(x)_f \psi(x)_g  = \langle \bar \psi(0) \psi(0) \rangle \left ( e^{i {\vec \pi \cdot \sigma\over 2 f_\pi}} \right )_{fg} 
  \end{equation}
  We only have to solve for $\pi_0$.
  
  A particularly simple case is that of $m_u = m_d$.  Then the potential is independent of $\pi_0$:
  \begin{equation}
  V(a) = -m_\pi^2 f_\pi^2 \cos({\cal N}{a \over f_a} ) = -{f_a^2 m_a^2 {\cal N}^{-2}}\cos({\cal N}{a \over f_a} ) .
  \end{equation}
  The presence of domains for ${\cal N} \ne \pm 1$ is manifest; the potential has ${\cal N}$ degenerate minimima with
  \begin{equation}
  {a \over f_a} = {2 \pi k \over {\cal N}}; k=0,\dots,{\cal N}-1,
  \end{equation}
  and the existence of domain wall solutions follows.
  The domain wall solution can be written down explicitly; it is the static soliton of the Sine-Gordan theory:
 \begin{equation}
 a(x) =  f_a\left  (4{\arctan (e^{m_a x}) \over {\cal N}} + {2 \pi k \over  {\cal N}} \right ).
 \end{equation} 

The tension of the domain wall satisfies:
\begin{equation}
T \propto f_a^2 m_a \propto {f_a m_\pi f_\pi}.
\end{equation}
It $m_u \ne m_d$, there is an additional contribution from the pion fields proportional to 
\begin{equation}
\delta T \propto {m_u -m_d \over m_u + m_d}{f_a m_\pi f_\pi}.
\end{equation}
So, in general, the pions make an order one contribution to the tension.  In the final collapse of the domains, this will be associated with production of energetic hadrons.
 
 \section{Domain Wall Cosmology}
 \label{cosmology}
 
 Domain walls, if they come to dominate the energy density of the universe, are problematic\cite{kolbturner}.  The domain wall energy density decreases as $1/R$, so it can quickly overwhelm the density of
 radiation or matter, falling as $1/R^4$ or $1/R^3$, respectively.  So it is necessary that there either never were domain walls at all, or that they disappear relatively quickly, typically by times of order a few seconds after the big bang.  Reference \cite{sarkar}
 considers, in addition to the two proposed solutions we mentioned earlier, a third possibility, that of a biased domain wall distribution, but rules it out.  They then argue
 that the constraints associated with PQ violating operators have been underestimated.  We will address their critique shortly.

 Reference \cite{sikiviedomainwalls} analyzed explicit PQ symmetry violation, tilting the axion potential
 and causing all but one type of domain to collapse.  Suppose the splitting between states is:
 \begin{equation}
 \Delta V = \epsilon ~10^{-10} ~m_\pi^2 f_\pi^2.
 \end{equation} 
 This corresponds to a potential for the axion roughly of the form:
 \begin{equation}
 \Delta V(a) = \epsilon~ 10^{-10} m_a^2 f_a~ a.
 \end{equation}
 $\epsilon$ cannot be extremely small if the domain wall system is not to dominate the energy of the universe before it disappears.
 
 When the temperature is of order $f_\pi$, the domain walls form.  The corresponding
  Hubble parameter is $H_0 = {m_\pi f_\pi \over M_p}$.  Calling the corresponding scale factor $R_0$, the domain wall density is subsequently of order:
  \begin{equation}
  \rho_{DW} = (f_a m_\pi f_\pi){ m_\pi f_\pi \over M_p} {R_0 \over R}.
  \end{equation}
  So domain walls dominate when
  \begin{equation}
  \left ({R_0 \over R}\right )^3 \approx {f_a \over M_p}
  \end{equation}
  or
  \begin{equation}
  \left ({R_0 \over R}\right ) \approx 10^{-2} \left ( {f_a\over 10^{12}} \right )^{1/3}.
  \end{equation}
  This corresponds to a temperature of order $1$ MeV, or times of order 10 seconds.  How quickly the domains collapse is the subject of the next section.
  
%  How quickly do the domain walls collapse?  We can ask, first, how long it takes them to reach the speed of light.  From
 % \begin{equation}
 % f_\pi m_\pi f_a \ddot x = \Delta V,
 % \end{equation}
%we have
%\begin{equation}
%\dot x \approx 10^{-10} \epsilon {m_\pi \over f_a} t
%\end{equation}
%or $v \sim 1$ for
%\begin{equation}
%\tau \sim 10^{22} \epsilon^{-1} {\rm GeV}^{-1}.
%\end{equation}
%Since the horizon size is order $t= \tau$ at this time, this is the time it takes the domain walls to dissappear.
%So $\epsilon$ cannot be very small.

\subsection{Fate of the Domain Walls}
\label{tilt}

As noted in the literature, when the walls collapse, their energy can be converted to kinetic energy of the domain walls, to axions, gravitational waves, electromagnetic radiation, and possibly other types of matter or radiation.  We will shortly
argue that, before collapse, the energy is principally converted to kinetic energy of the walls, followed by highly relativistic axions.  At the final collapse, this energy is converted to extremely relativstic
axions.    Gravitational and electormagnetic radiation are minor components of the energy budget.  These axions would still be highly relativistic today.

When formed, the bubbles have radius, $r_0$, of order:
\begin{equation}
r_0 \approx {M_p \over  m_\pi f_\pi}.
\end{equation}
Initially their acceleration due to $\Delta V$ is slightly less than that due to the Hubble expansion, $H$:
\begin{equation}
a \sim {\Delta V \over f_a f_\pi^2} \approx \epsilon  10^{-22} f_\pi^2;~~H \sim 10^{-19} f_\pi^2.
\end{equation}
They become comparable when the temperature decreases by a factor of order $10^2$.  beyond this point we can, to first approximation, neglect the expansion.  The velocity becomes of order one in less than a Hubble time, and collapse occurs in such a time.  We can then ask:  if we ignore gravitational radiation, what is the velocity of the domain wall once the domain shrinks to a microscopic size (more precisely, how large is the Lorentz $\gamma$ factor for the wall.  

We take as the initial time the time when the cosmic acceleration is equal to the acceleration of the wall:
\begin{equation}
H_0 = {\Delta V \over f_\pi^2 f_a} \approx \epsilon \times 10^{-24} ~{\rm GeV}.
\end{equation}
The energy stored in a horizon sized region of one of the excited states is:
\begin{equation}
E \approx \Delta V H_0^{-3}  \sim \epsilon 10^{58}~ {\rm GeV}.
\end{equation}
If there is no emission of axions, gravitational or other radiation during the collapse, then once the region size is of order $m_a^{-1}$, the $\gamma$ factor is enormous.  The effective mass of the system is of order
\begin{equation}
m_{eff}=  f_a f_\pi^2 m_a^{-2} \sim 10^{38}
~{\rm GeV},
\end{equation}
so
\begin{equation}
\gamma \sim \epsilon 10^{20}.
\end{equation}

%Actually, for the problem of domain collapse, one should allow for the expansion which occurs initially, as described above.  So $\gamma$ at collapse is actually several orders of magnitude larger.  We will use the estimate above as a more ``conservative" one.

We now argue that in fact most of the energy gained is transferred to kinetic energy of the domain wall.
Initially the domain is large and the curvature of the domain wall is negligible on macroscopic scales.
To  consider axion radiation, it is helpful to work in the instantaneous rest frame of the domain wall (more precisely, of a macroscopic segment of the wall).  We can define what we mean as axion radiation
by considering a set of domain walls, instantaneously at rest, described by a classical field configuration,
\begin{equation}
\phi(\vec x) = \phi_{cl}(\vec x-\vec x_0).
\end{equation}
Were it not for the symmetry-breaking potential, $\Delta V$, these configurations would be solutions of the equations of motion if $\vec x_0 = \vec v t$.  But due to the potential, they are not.  
Axion radiation corresponds roughly to the difference, $\delta \phi$, of the actual axion field and the would-be domain wall configuration at a given time.  This is proportional to 
\begin{equation}
\delta \phi(\vec x,t) = {\rm C} \ddot  x_0(t)_i \cdot \partial_i \phi_{cl} (\vec x,t). 
\end{equation}
It would be challenging to compute the energy carried by $\delta \phi$, but it's form, for non-relativistic motion, can be determined by simple considerations.  The energy should be rotationally invariant,
translationally invariant, and time reversal invariant (up to very small effects within the Standard Model coupled to an axion).  So the energy per unit time transferred to $\delta \phi$ behaves as:
\begin{equation}
{\cal E} = {\rm B} f_a^2 \vert \ddot x_0^2 \vert A
\end{equation}
where $A$ is the area of the domain wall, and ${\rm B}$ is an order one constant.  We can write $\ddot x_0$ in terms of $\Delta V$  and the tension of the domain wall,
\begin{equation}
\ddot  x_0 = {\Delta V \over f_\pi m_\pi f_a}.
\end{equation}
Correspondingly, in a frame boosted with Lorentz factor $\gamma$, the energy per unit time is increased by a factor of $\gamma$ for the energy, but decreased by a factor of order $\gamma$ from the time dilation,
so the transformation between the two frames behaves as $(\gamma)^0$.

A little more precisely, we might consider radiation in a time interval $\Delta t$ in the instantaneous rest frame of the wall.  $\Delta t$ should be such that the wall is non-relativistic in that frame.  For the element
of area $A$, there will be movement of order $\Delta t \times v$ in this time period. The corresponding time elapsed in our observers frame is:
\begin{equation}
\Delta t^\prime = \gamma (\Delta t +v \Delta z)
\end{equation}
but the second term is, by assumption, much smaller than the first and can be neglected.

We want to compare the energy radiated by the wall per unit time with the energy the wall acquires per unit time from $\Delta V$.  This is 
\begin{equation}
\Delta V A,
\end{equation}
so the condition that radiation is comparable to the energy increase of the wall per unit time due to $\Delta V$ is
\begin{equation}
{\Delta V \over f_\pi^2 m_\pi^2} >1
\end{equation}
which is never satisfied.  In other words, the energy radiated in axions as the walls collapse is negligible.  Similar considerations lead to suppression of electromagnetic and gravitational
radiation. So the walls are extremely relativistic when they finally shrink to microscopic size.  At this point, one expects that the collapse results in production of extremely relativistic axions,
with $\gamma$ factors comparable to those we discussed above, and very relativistic hadrons.  The hadrons quickly thermalize with the background hadrons.  The number of axions produced would be small, consistent with the fact that the final domains are microscopic in size.  
As we explain below, very little of the axion energy would be degraded in collisions with hadrons. In any case, provided that the domain walls don't
dominate the energy density of the universe at the time of their collapse, their cosmological effects would be minor.

\subsection{The Final Stage of Domain Wall Collapse}

Because of the enormous $\gamma$ factors of the axion produced in the decay of a domain, most of these axions stream through the universe.  Only rarely does one interact with
quarks, gluons. or other axions.  For $s$-channel processes, the mean time between collisions is long.  Even assuming an order one coupling of these high energy axions to nucleons,
\begin{equation}
{\delta {\cal L}} = a \bar N \gamma_5 N,
\end{equation} 
the mean free time for axion collisions with nucleons is
\begin{equation}
\tau \sim 10^{10} \gamma m_a T^{-3}m_N \sim 10^{33} \left ({\gamma \over 10^{20}} \right ) \left ({{\rm MeV} \over T} \right )^3 \left ({m_a \over 10^{-14}} \right )                           
\end{equation}
in ${\rm GeV}$ units, and we have taken ${n_B \over n_\gamma} \approx 10^{-10}$.  This is an enormous time, comparable to even the current age of the universe.  As we have noted, the $\gamma$
factor is actually likely to be serval orders of magnitude larger than $10^{25}$, giving further suppression.

For $t$-channel exchange, due to the enormous $\gamma$ factor, scattering is appreciable only at extremely small angles.  As a result, the mean scattering time is, again, extremely large.
So most of these axions are still around today, and are highly energetic.  Their numbers, however, are not large, and their contribution to the energy budget of the present universe is smaller
than the photon contribution (assuming that the domain wall contribution was always a small fraction of the total energy density at the time of collapse).  Interactions with ordinary matter are very rare.

\section{Anthropic Considerations}
\label{anthropic}

As we have stressed, there is one troubling feature of the Peccei-Quinn symmetry for these relatively low $f_a$ axions in the suppression required for a Peccei-Quinn symmetry
of sufficient quality to solve the strong CP problem, and another in the requirement of sufficient tilt solution to solve the domain wall problem.  One might, indeed, argue
that the requirements to obtain a PQ symmetry of sufficient quality to solve the strong CP problem are implausible,  For example, if the result of a discrete symmetry, the symmetry must be very large\cite{kmr,dinediscretesymmetries}.
On the other hand, we have seen that to avoid a domain-wall dominated universe, the tilt must be just barely smaller than that required for the PQ symmetry.  One might be inclined to discard axions with
these relatively low decay constants, but
it is also tempting to ask whether such bizarre constraints might be satisfied as a result of anthropic considerations.  In this section, we will examine this possibility.  We do not attempt
to understand precisely how, within some sort of landscape, this might be realized in detail.  Rather we ask the simpler question:  might the existence of observers be ruled out if we made a change
in a single parameter, holding others fixed.

If the tilt is controlled by a discrete symmetry, we might imagine that the symmetry violating terms in the axion potential have the form $A M_p^{-n}\Phi^{n+4}$, where $\Phi =  f_a e^{i {a \over f_a}}$, $A$ is now a complex constant ({\it not the area}) and
$\Phi \rightarrow e^{{2 \pi i \over N}} \Phi$ under the $Z_N$ symmetry, so 
\begin{equation}
\delta V = \vert A\vert f_a^4 \left ( {f_a \over M_p} \right )^{N-4} \cos((N{a \over f_a} + \alpha).
\end{equation}
In order that $\theta$ be small enough, assuming $A$ is of order one, $N$ must be quite large.  If $f_a = 10^{12}$ GeV,  we require $N \ge 13$, for example.  At the same time, our discussion
of domain wall evolution implies that $N$ can't be larger than $13$.

The lower limit on the tilt (upper limit on $N$) is a relatively easy one to explain anthropically.  Were domain walls to dominate the energy density of the universe before their collapse, the universe would not evolve to a situation with structures of the sort
we see in nature (and which support observers). 

The upper limit might be understood if we assume that a dark matter density close to that observed is necessary for (or perhaps optimizes the number of) observers.  Write the tilt contribution to the axion
potential as
\begin{equation}
\delta V(a) = \epsilon 10^{-10} m_\pi^2 f_\pi^2 \cos({a \over f_a} + \alpha).
\end{equation}
We require $\epsilon < 1$.  Suppose $\epsilon$ was much larger, corresponding to $N=11$, for example, and $\epsilon \sim 10^4$.  Then the axion mass receives a larger contribution
from $\delta V$ then from QCD, and it starts to oscillate when the temperature is of order $10^5$ GeV.  As a
result, there are far too few axions to constitute the dark matter.   Axions only account for about $10^{-20}$ of the energy density initially.   At temperatures of order $1$ eV, they are only $10^{-6}$ of
the total energy of the universe.

The case of $n=12$ is different.  The contribution, at zero temperature, to the axion mass from $\delta V$ is smaller than that from QCD, but we have to investigate the behavior of the
system as a function of temperature.  The axion mass as a function of temperature behaves as\cite{kolbturner}:
\begin{equation}
m_a(T) \approx m_a \left ({\Lambda_{QCD} \over T} \right )^{3.7}.
\label{axionmasstemperature}
\end{equation}  
Without $\delta V$, the axion begins to oscillate when 
\begin{equation}
3 H = m_a(T).
\end{equation}
This corresponds to a temperature of order $10$ GeV.  But $m_a(T) < \delta m_a$ down to a smaller temperature, of order $0.2$ GeV if we take
the formula \ref{axionmasstemperature} seriously at these temperatures. So there is significant overproduction of dark matter.

All of this is meant to establish that there is a plausible anthropic rationale for the size of $\delta V$ being such that axions constitute the dark matter, and the domain wall system collapses ``just in time".
We stress again that we don't have a detailed cosmological picture for how this might be implemented.

  \section{Conclusions}
  \label{conclusions}
  
  In this paper we have adopted
  the conventional view of axion cosmology, that the PQ transition occurred after inflation and that the post inflationary universe was once at a temperature well above
  the scales of QCD, and focused on the resulting question of domain walls.  We have recalled that the problem is generic, and admits only a small number of solutions.  Among these, we
  have considered the effects of small, explicit breaking of the symmetry.  We have recalled the well-known issue that the constraints of obtaining small enough $\theta$ and domain wall
  annihilation before domain wall dominance restrict the symmetry breaking to a narrow range.  We have pointed out that it would be almost absurd for the size of this breaking to be a consequence of discrete
  symmetries.  We have noted that, as for other seemingly absurd phenomena actually observed in nature, one might contemplate an anthropic solution.  This has the virtue that it could explain the remarkable
  coincidences required. 
  
  But most of our attention has been devoted to the fate of the universe in such a picture.  We have studied the question of where the energy in the domain walls goes.  We have noted that
  gravitational radiation leads to a limiting Lorentz factor, $\gamma$, which in turn implies that the vast majority of the domain wall energy is converted to gravitational waves.  This is in contrast
  to the possibility that much of the energy ends up in non-relativistic, dark matter axions\cite{sarkar}.  The resulting constraints are mild, in the sense that they are not much stronger than the requirement
  that the domain wall energy not dominate the energy density of the universe before annihilation.
    
\section*{Acknowledgments}
We thank Patrick Draper, Guido Festuccia and Pierre Sikivie for conversations and critical comments.
This work was supported in part by U.S. Department of Energy grant No. DE-FG02-04ER41286.

\bibliography{domain_walls}
\bibliographystyle{JHEP}

\end{document}